\documentclass[12pt]{article}

\usepackage[utf8]{inputenc}
\usepackage[english]{babel}
\usepackage{helvet}
\usepackage[centertags]{amsmath}
\usepackage{amssymb}
\usepackage{bm}
\usepackage{amsbsy}
\usepackage{graphicx}
\usepackage{appendix}
\usepackage{mathrsfs}
\usepackage{epsfig}
\usepackage{color}
\usepackage{euscript}
\usepackage{ulem}
\usepackage{multirow}
\usepackage{cite}
\newcommand{\mbC}{\mathbb C}
\newcommand{\mbR}{\mathbb R}
\newcommand{\mbP}{\mathbb P}

\newcommand{\fG}{\mathfrak{G}}

\newcommand{\cH}{\mathcal{H}}
\newcommand{\cI}{\mathcal{I}}
\newcommand{\cE}{\mathcal{E}}
\newcommand{\cJ}{\mathcal{J}}
\newcommand{\cK}{\mathcal{K}}
\newcommand{\cL}{\mathcal{L}}

\newcommand{\cS}{\mathcal{S}}
\newcommand{\cP}{\mathcal{P}}
\newcommand{\cV}{\mathcal{V}}

\newcommand{\cU}{\mathcal{U}}

\newcommand{\cZ}{\mathcal{Z}}
\newcommand{\rT}{\mbox{\footnotesize T}}

\newcommand{\fP}{\mathfrak{P}}

\newcommand{\fN}{\mathfrak{N}}

\newcommand{\fM}{\mathfrak{M}}

\newcommand{\vqp}{\begin{pmatrix}
    q\\p
\end{pmatrix}}
\newcommand{\rqs}{\begin{pmatrix}
    r\\s
\end{pmatrix}}
\newcommand{\vqpf}{\begin{pmatrix}
    q_1\\p_1
\end{pmatrix}}
\newcommand{\vqps}{\begin{pmatrix}
    q_2\\p_2
\end{pmatrix}}

\definecolor{dgreen}{rgb}{0,0.6,0}

\definecolor{darkblue}{rgb}{0., 0, 1}

\definecolor{purple}{rgb}{0.65,0.,0.78}

\definecolor{orange}{rgb}{0.89,0.42,0.05}

\usepackage{jheppubm}

\newcommand{\uC}{\underset{\mbC}{\otimes}}
\newcommand{\uR}{\underset{\mbR}{\otimes}}

\newcommand{\ucK}{\underset{\cal K}{\otimes}}

\newcommand{\be}{\begin{equation}}
\newcommand{\ee}{\end{equation}}
\newcommand{\bea}{\begin{eqnarray}}
\newcommand{\eea}{\end{eqnarray}}
\newcommand{\nn}{\nonumber}
\numberwithin{equation}{section}

\title{Notes on Real  Quantum Mechanics in a Kähler Space
}

\author{Irina Aref'eva and Igor Volovich}
\affiliation{Steklov Mathematical Institute, Russian Academy of Sciences, \\ Gubkina Street, 8, 119991 Moscow, Russia}
\emailAdd{\\
arefeva@mi-ras.ru}
\emailAdd{\\
volovich@mi-ras.ru}

\abstract{The necessity of complex numbers in quantum mechanics has long been debated. This paper develops a real Kähler space formulation of quantum mechanics \cite{Volovich:2025rmi}, asserting equivalence to the standard complex Hilbert space framework. By mapping the complex Hilbert space $\mbC^n$
to a real Kähler  space $\cK ^{2n}$, i.e. 
$\mbR^{2n}$
equipped with a metric, a symplectic structure 
 and an automorphism, we establish a correspondence between Hermitian operators in $\mbC^n$
 and real operators in $\cK^{2n}$. 
  While the isomorphism appears straightforward 
 some subtleties emerge: 
 (i) the overcounting of composite system states under tensor products in $\mbR^{2n}$, and (ii) the double degeneracy of operator spectra in the real formulation. Through a systematic investigation of these challenges, we clarify the structural relationship between real and complex formulations, resolve ambiguities in composite system representations, and analyze spectral consequences. Our results reaffirm the equivalence of the two frameworks while highlighting nuanced distinctions with implications for foundational debates on locality, phase invariance, and the role of complex numbers in quantum theory.
}

\begin{document}

\maketitle
\section{Introduction}\label{Sect:intro}

In quantum mechanics complex numbers play a very important role.  The question of why quantum theory requires complex numbers has been a subject of debate for many years \cite{Stueckelberg,BVN,Varadarajan,Arvind:1995ab,
Zhu:2020iml,Renou:2021dvp,Chen:2021ril, Li:2021uof,Bednorz:2022zgq,2103.12740,Vedral:2023pij,2407.12755,Takatsuka:2025hez,Sarkar:2025tmd,VEBEQ,Hita:2025okv,Hoffreumon:2025nmq,
Feng:2025eci,Volovich:2025rmi} and refs therein.
\\

 In the previous paper of one of us \cite{Volovich:2025rmi}, it was  proposed a formulation of real quantum mechanics in a real Kähler space, which as it was claimed is equivalent to the standard formulation of quantum mechanics in a complex Hilbert space. Therefore, all results obtained within the framework of the real Kähler space approach will be equivalent to the corresponding results of standard quantum mechanics in the complex Hilbert space approach, including the principle of locality. Moreover there it was also claimed  that quantum mechanics, usually formulated in the Hilbert space over complex numbers \cite{Dirac,Varadarajan,Landau:1991wop}, can be reformulated in the Kähler space over real numbers. Main formula behind this claims are the following: 
\\
 
To a finite-dimensional complex Hilbert space $\mbC^n$ with inner product $\langle \psi_1,\psi_2\rangle $, \\$\psi_i \in \mbC^n,i=1,2,$ one can associate a real Kähler space $\cK^{2n}$ consisting of pairs $\begin{pmatrix}
    q\\p
\end{pmatrix}$, constructing from real and imaginary parts of $\psi=q+i p$ and forming a real Hilbert space ${\mbR}^{2n}$ with inner product $g\left(\vqpf,\vqps\right)=\Re(\langle \psi_1,\psi_2\rangle)$, 
which is also equipped with a symplectic form $\omega$ defined as $\omega\left(\vqpf,\vqps\right)=\Im (\langle \psi_1,\psi_2\rangle)$ and an automorphism $J\vqp=\begin{pmatrix}
    -p\\q
\end{pmatrix}$ \cite{Volovich:2025rmi}.
Therefore, there is a following relation between the  complex Hilbert space and the real Kähler space
\be\label{eq:inner-INT}   \langle \psi_1, \psi_2\rangle = g\left(\vqpf,\vqps\right)+ 
i \omega\left(\vqpf,\vqps\right), \quad \psi_i=q_i+ip_i,\quad i=1,2.
\ee
Therefore, there is a natural isomorphism between  operators in the Hilbert $\mbC^n$ space and in the Kähler space $\cK ^{2n}$.
At first glance, the connection  \eqref{eq:inner-INT}  suggests equivalent formulations of quantum mechanics in Hilbert and Kähler spaces. However, several subtleties arise:
i) The first subtlety concerns the treatment of quantum states. In standard formulations, states correspond to equivalence classes of vectors distinguished only by multiplication by a phase factor.
ii) The second subtlety involves defining states for composite systems. Specifically, naively interpreting the tensor product in Kähler space as a tensor product in  $\mbR^{2n}$ leads to an overcounting of composite system states in the real Kähler space compared to the conventional Hilbert space formulation.
iii) The third subtlety concerns the spectrum of the Hermitian Kähler operator $\cL$ compared to the spectrum of the corresponding Hermitian operator $L$: it is doubly degenerate relative to the spectrum of the corresponding Hermitian operator 
$L$.
The purpose of this paper is to systematically study these subtleties.
\\

By reformulating quantum mechanics in the real Kähler space \(\mathcal{K}^{2n}\), we preserve all essential features of the theory, including the consistent treatment of composite systems via tensor products and adherence to the standard postulates. Crucially, we resolve ambiguities arising in two key areas: 
1) The avoidance of overcounting in composite systems through careful treatment of tensor products in \(\mathbb{R}^{2n}\).
2) The interpretation of spectral degeneracy in real operator representations.
\\

Let us highlight  connections to previous papers. The geometric formulation of quantum mechanics on {\it Kähler manifolds} discussed in \cite{Ashtekar:1997ud,Kible} differs from our approach on {\it linear Kähler spaces}. For example, in the Appendix, section \ref{Cn-K2n}, we deal with $\cK^2$ rather than with projective manifolds such as \(\mathbb{C}P^1\). This distinction avoids the nonlinear complexities of curved spaces while retaining symplectic and metric structures.
\\

Regarding debates on real versus complex formulations, recent works present conflicting perspectives on the necessity of complex numbers in quantum theory:

1) \textit{Pro-complex arguments} are advanced in \cite{Renou:2021dvp} and \cite{Sarkar:2025tmd}. These papers claim that real-number quantum theories cannot replicate complex-number predictions in multipartite network scenarios, and propose experimental tests to falsify real formulations.

2) \textit{Real-formulation viability} is defended in \cite{Hita:2025okv,Hoffreumon:2025nmq,Volovich:2025rmi}, which claim that real quantum mechanics can reproduce all multipartite predictions when composite systems are properly defined. This perspective frames complex numbers as a convenience rather than a necessity.  In this work, we adopt the latter perspective.
\\

The paper is organized as follows. 
In Sect. \ref{KS}, we provide a formal definition of real finite-dimensional Kähler spaces.
In Sect. \ref{RHK}, we discuss relationships between Hilbert and Kähler spaces.
In Sect. \ref{QMK}, we present the formulation of quantum mechanics in real Kähler spaces, paying special attention to spectral decomposition and the wave functions of composite systems.
Finally, in Sect. \ref{QMKH}, we compare quantum mechanics in $\mbC^n$
 and 
$\cK^{2n}$ spaces.
  The paper ends with an appendix. In Appendix 1,  proofs of some relationships from the main text are given, and Appendix B follows with specific examples of Kähler spaces $\cK ^2$ and $\cK ^4$.

\section{Kähler  Spaces}\label{KS}  
Here and thereby we consider a finite dimensional Kähler space.
\subsection{Finite dimensional Kähler space}
A  finite dimensional Kähler space is a real finite dimensional Hilbert space equipped with a symplectic form and an automorphism called complex structure. 
A Kähler space is a quadruplet 
$\cK^{2n}=(\mbR^{2n}, g,\omega, J)\), where:  
\begin{itemize}  
    \item $g: \mbR^{2n} \times \mbR^{2n}\to \mathbb{R}$ is a positive defined 
      symmetric, non-degenerate bilinear form, 
  
    \item $\omega: \mbR^{2n} \times \mbR^{2n} \to \mathbb{R}$ is a skew-symmetric, non-degenerate bilinear form (i.e., a symplectic form),  
    \item $J: \mbR^{2n} \to \mbR^{2n}$ is an automorphism satisfying $J^2 = -\mathrm{id}$ (referred to as a complex structure),
  \item  the following relations hold 
  \bea \label{g-omega}g(\cdot, \cdot)&=&\omega(\, \cdot\,, J\,\cdot\,), \\
  \label{omega-g}\omega(\, \cdot\,, \,\cdot\,)&=&g(J\, \cdot\,, \cdot\,), \\
  \label{omegagJ}
  \omega(J\, \cdot\,,J \cdot\,)&=&\omega(\, \cdot\,, \cdot\,); 
  \eea
\end{itemize}  
 see, for example, \cite{Kahler,LKL,Borowiec:2000af}.

\subsection{Kähler structure in $\mbR ^{2n}$}
Since $\mbR ^{2n}=\mbR ^{n}\oplus \mbR ^{n}$
and we represent elements of $\mbR ^{2n}$ as pairs \be\label{qpn}
\begin{pmatrix}
    q\\p
\end{pmatrix}\in \mbR ^{2n},\quad q,p\in \mbR^n
.\ee
Vectors $p$ and $q$  in \eqref{qpn} can be expand into bases as 
\be\label{qp}
q =\sum _{a=1}^n q_a e_{a}, \quad q_a\in \mbR,
\qquad p=\sum _{a=1}^n p_a h_{a}, \quad p_a\in \mbR,\ee
For simplicity, we can take the same bases in both $\mbR ^{n}$, i.e. $e_a=h_{a}$.
\\

Now we introduce the Kähler structure in 
$\mbR ^{2n}$:
\begin{itemize}\item
i) the scalar product on $\mbR ^{2n}$
we define as 
\be \label{com-g}
g\left(\begin{pmatrix}
  q_1\\  p_1
\end{pmatrix},\begin{pmatrix}
  q_2\\  p_2
\end{pmatrix}\right)=\sum _{a=1}^N\, (q_{1,a}q_{2,a}+p_{1,a}p_{2,a})\ee
\item
ii) the  symplectic form on $\mbR ^{2N}$
we define as
\be \label{com-omega}
\omega\left(\begin{pmatrix}
  q_1\\  p_1
\end{pmatrix},\begin{pmatrix}
  q_2\\  p_2
\end{pmatrix}\right)=\sum _{a=1}^N\, (q_{1,a}p_{2,a}-q_{2,a}p_{1,a})\ee
\item iii) the complex  structure we define 
as \be
\label{mCS}
\cJ=\begin{pmatrix}
  0&-I\\I&0  
\end{pmatrix} \quad \mbox{and} 
\quad\cJ\begin{pmatrix}
  q\\  p
\end{pmatrix}=\begin{pmatrix}
  -p\\  q
\end{pmatrix},\ee
where $I$  is the $n\times n$ identity matrix. 
\end{itemize}

One can check that relations 
\eqref{g-omega}, \eqref{omega-g}
and \eqref{omegagJ} hold.
\subsection{Operators}
In this subsection, we consider linear operators on the Kähler space  $({\mathbb R}^{2n},\omega,\cJ,g)$. 

Since elements of $\cK$ are represented as pairs of vectors \eqref{qp} linear operators  ${\cal L}$ on $\cK$ can be represented   as  block matricies
\be
\cL = \begin{pmatrix} S_1 & A_1 \\ A_2 & S_2 \end{pmatrix}\quad \mbox{and}
\quad  \cL \begin{pmatrix} q \\ p \end{pmatrix}= 
\begin{pmatrix} S_1q + A_1p \\ A_2q + S_2p \end{pmatrix}.
\ee

\subsubsection{Double degeneration of spectrum of $\cK$-Hermitian operators}
\label{Res-of-unity}

We define a $\cK$-Hermitian operator in the Kähler space $(\mathbb{R}^{2n}, \omega, \cJ, g)$ as a real $2n\times 2n$ matrix $\cL$ that satisfies the conditions
\begin{equation}\label{cLJ}
    \cL^{\rT} = \cL, \quad \cL \cJ = \cJ \cL,  
\end{equation}
where $\cL^{\rT}_{ij} = \cL_{ji}$ and $\cJ$ is defined as in \eqref{mCS}. Matrices satisfying \eqref{cLJ} admit the block decomposition:
\begin{equation}\label{cL}
\cL = \begin{pmatrix} 
S & -A \\ 
A & S 
\end{pmatrix},
\end{equation}
where 
\begin{align}
A^{\rT} &= -A, \label{eq:A_transpose} \\
S^{\rT} &= S. \label{eq:S_symmetric}
\end{align}

To demonstrate spectral degeneracy, consider the eigenvalue problem for $\cL$:
\begin{equation}\label{eigenstate2c}
(\cL - \lambda I) \begin{pmatrix}
  q_\lambda \\ p_\lambda  
\end{pmatrix} = 0.
\end{equation}
Since $\cL$ satisfies \eqref{cLJ}, multiplying \eqref{eigenstate2c} by $\cJ$ yields
\begin{equation}\label{eigenstate2cm}
(\cL - \lambda I) \, \cJ \begin{pmatrix}
  q_\lambda \\ p_\lambda  
\end{pmatrix} = 0,
\end{equation}
revealing a new eigenvector $\cJ \begin{pmatrix} q_\lambda \\ p_\lambda \end{pmatrix}$ with identical eigenvalue $\lambda$. Thus, the spectrum of any $\cK$-Hermitian operator $\cL$ is doubly degenerate.

The $\cK$-Hermitian operators in $\mathbb{R}^{2n}$ admit the following spectral representation (resolution of the identity):
\begin{align}
\sum_{i=1}^{n} \lambda_i (\cP_{1i} + \cP_{2i}) &= \cL, \label{PL} \\
\sum_{i=1}^{n} (\cP_{1i} + \cP_{2i}) &= \cI, \label{PI}
\end{align}
where $\cI$ is the $2n\times 2n$ identity matrix, and $\cP_{\alpha i}$ ($i = 1, \ldots, n$, $\alpha = 1, 2$) are projectors with block form
\begin{equation}
\cP_{1i} = \begin{pmatrix}
P_i & 0 \\
0 & 0
\end{pmatrix}, \quad 
\cP_{2i} = \begin{pmatrix}
0 & 0 \\
0 & P_i
\end{pmatrix}, \quad 
P_i^2 = P_i,
\end{equation}
satisfying the conditions:
\begin{align}
\cP_{\alpha i}^2 &= \cP_{\alpha i}, \quad 
\cP_{1i} \cJ = \cJ \cP_{2i}, \label{eq:proj_cond1} \\
\cP_{1i} \cP_{2i} &= 0, \quad 
\cP_{\alpha i} \cP_{\beta j} = 0 \quad (\alpha \neq \beta). \nonumber
\end{align}

Equations \eqref{PL} and \eqref{PI} can be rewritten as
\begin{align}
\sum_{i=1}^{n} \lambda_i \cE_{i} &= \cL, \label{EL} \\
\sum_{i=1}^{n} \cE_i &= \cI, \label{EI}
\end{align}
where $\cE_i$ relates to $\cP_{1i}$ and $\cP_{2i}$ through
\begin{equation}
\cE_i = \cP_{1i} + \cP_{2i},
\end{equation}
and satisfies the properties:
\begin{equation}
\cE_i^2 = \cE_i, \quad 
\cE_j \cE_i = 0 \quad (j \neq i).
\end{equation}

 \subsubsection{Kähler unitary group $\cU(n)$, symplectic group $Sp(2n, \mathbb{R})$, and orthogonal group $O(2n)$}\label{Kunitary}

A Kähler unitary group $\cU(n)$ consists of matrices $\cU$ in the form
\begin{equation}
\label{cV}
  \cU = 
  \begin{pmatrix}
    X & Y \\
    -Y & X
  \end{pmatrix},
\end{equation}
where $X, Y$ are real $n \times n$ matrices, and $\cU$ matrices are subject to the relation
\begin{equation} \label{cVXY}
  \cU \cU^{\rT}  = \cI.
\end{equation}
Relations \eqref{cVXY} give the following conditions:
\be
  X X^{\rT} + Y Y^{\rT} = X^{\rT} X + Y^{\rT} Y = I, 
\ee
The matrices in the form \eqref{cV} are also elements of $O(2n, \mathbb{R})$, i.e., $\cV \in O(2n, \mathbb{R})$. Moreover, they are elements of the symplectic group $Sp(2n, \mathbb{R})$.
Indeed, a symplectic matrix $\cS$ of the symplectic group $Sp(2n, \mathbb{R})$ satisfies the relation
\begin{equation}
  \cS^{\rT} \cJ \cS = \cJ,
\end{equation}
where $\cJ$ is given by \eqref{mCS}. Representing the $2n \times 2n$ matrix $\cS$ in the block form
\begin{equation*}
  \cS = \begin{pmatrix}
    A & B \\
    C & D
  \end{pmatrix},
\end{equation*}
where $A, B, C, D$ are $n \times n$ matrices, we obtain the following conditions on $A, B, C$, and $D$:
\bea
&&A^{\rT} C, \quad B^{\rT} D, \quad A B^{\rT}\, \mbox{and}\quad  C D^{\rT}\quad \mbox{ are symmetric}, \\
&&A^{\rT} D - C^{\rT} B = I,\quad A D^{\mathrm{T}} - B C^{\mathrm{T}} = I. 
\eea

If we assume $A = X = D$ and $B = -C = Y$, as in \eqref{cV}, the listed conditions reduce to:
\bea
 Y^{\rT} X = X^{\rT} Y,\quad  X Y^{\rT} = Y X^{\rT},\\  X^{\rT} X + Y^{\rT} Y = I,\quad 
  X  X^{\rT}+ Y Y^{\rT}  = I.
 \eea

This implies the fundamental relation:
\begin{equation}
  U(n) = \mathrm{Sp}(2n, \mathbb{R}) \cap \mathrm{O}(2n, \mathbb{R}),
\end{equation}
where $U(n)$ is the unitary group, $\mathrm{Sp}(2n, \mathbb{R})$ is the symplectic group, and $\mathrm{O}(2n, \mathbb{R})$ is the orthogonal group.

\subsection{Tensor products on Kahler spaces}\label{TensorProd}
In this subsection we consider two natural definitions of tensor products of real Kähler spaces. We will elaborate explicit constructors using bases in corresponding spaces.

   In $\mbR ^{2n}=\mbR ^{n}\oplus \mbR ^{n}$ there is a basis, 
   \be \{\ell_{A},A=1,...2n\}=\{e_{a}|+\rangle,\,h_{b}|-\rangle, a, b=1,...n\},
   \ee
   where  $|+\rangle=\begin{pmatrix}1\\0
   \end{pmatrix}$, $e_{a}|+\rangle=\begin{pmatrix}e_{a}\\0
   \end{pmatrix}$, and $|-\rangle=\begin{pmatrix}0\\1
   \end{pmatrix}$, $h_{b}|-\rangle=\begin{pmatrix}0\\h_{b}
   \end{pmatrix}$ ;
   as noted above we take $h_a=e_a$.
   
To define the tensor product of two linear vector spaces, it is sufficient to define the tensor product on the elements of the bases. 

\subsubsection{
 The tensor product operator 
 $\uR $ on Kähler spaces}
 We can write
\bea 
\ell_A \otimes \ell_B&=&e_{a}\otimes e_{b}\Big[\sum_{i,j=1,2}|i\rangle \otimes |j\rangle \Big]\nn\\
  &=&e_{a}\otimes e_{b}\Big[\,|+\rangle \otimes|+\rangle
  +|+\rangle \otimes|-\rangle
  +|-\rangle \otimes|+\rangle
  +|-\rangle \otimes|-\rangle
  \Big]\label{tpR}\eea

 Denoting 
\be\label{eta12}
\begin{pmatrix}
   q_1\\p_1  
 \end{pmatrix}=\eta _1;\quad 
 \begin{pmatrix}
   q_2\\p_2  
 \end{pmatrix}=\eta _2,
\ee 
 we consider the tensor product in the sense of \eqref{tpR} that is indicated as $\uR$ and we get
 \bea\label{T12R}
 && \qquad\qquad\qquad\eta _1 
 \uR \eta_2=\begin{pmatrix}
  q_1\\  p_{1}
\end{pmatrix} \uR\begin{pmatrix}
  q_2\\  p_{2}
\end{pmatrix}\\\nn&&\qquad \quad =\sum_a \left(q_{1a} e_{a}|+\rangle+
 p_{1a} e_{a}|-\rangle\right)\uR\sum_b \left(q_{2b} e_{b}|+\rangle+
 p_{2b} e_{b}|-\rangle\right)\\\nn&=&
 \sum_{a,b} \Big(q_{1a} q_{2b}\,|+\rangle \otimes|+\rangle
  +
 p_{1a} q_{2b} 
 |-\rangle\otimes |+\rangle 
+q_{1a} p_{2b}\, |+\rangle \otimes|-\rangle
  +
 p_{1a} p_{2b} 
 |-\rangle\otimes |-\rangle \Big) e_{a}\otimes e_{b}.
 \eea

The  scalar product and the symplectic form of two vectors in the form \eqref{T12R}  are given by the formula 
 \bea
\label{TRg}g\left( \eta _1 \uR \eta_2, \chi _1 \uR \chi_2\right)&=&g\left( \eta _1 ,\chi _1 )\cdot g(\eta_2, \chi_2\right)\\
\label{TRo}
\omega\left( \eta _1 \uR \eta_2, \chi _1 \uR \chi_2\right)&=&
\omega\left( \eta _1 ,\chi _1 )\cdot g(\eta_2, \chi_2\right)+g\left( \eta _1 ,\chi _1 )\cdot \omega(\eta_2, \chi_2\right).
  \eea
   
\subsubsection{Tensor product operator $\ucK $ on
 Kahler Spaces}
 In \cite{Volovich:2025rmi} also has been introduced the $\cK$ tensor product
 \bea\nn
   && \qquad \qquad\eta _1 \ucK \eta_2=\begin{pmatrix}
  q_1\\  p_{1}
\end{pmatrix} \ucK\begin{pmatrix}
  q_2\\  p_{2}
\end{pmatrix}\\\nn
 &=&\sum_a \begin{pmatrix}
  q_{1a}\\  p_{1a}
\end{pmatrix} e_{a}\ucK\sum_b \begin{pmatrix}
  q_{2a}\\  p_{2a}
\end{pmatrix} e_{b}
=\sum_{a,b}\begin{pmatrix}
  q_{1a} q_{2b} - p_{1a} p_{2b}\\ q_{1a} p_{2b} +p_{1a} q_{2b}
\end{pmatrix} e_{a}\otimes e_{b} \\\label{T12C}
 &=&\sum_{a,b}
 \Big( (q_{1a} q_{2b} - p_{1a} p_{2b})|+\rangle+( q_{1a} p_{2b} +p_{1a} q_{2b})|-\rangle\Big)
 e_{a}\otimes e_{b}. 
\eea
This definition is in agreement with the tensor in $\mbC$, see \eqref{1ten} below.
Comparing \eqref{T12R} and \eqref{T12C} we get
\bea
\label{T12CR}
   && \eta _1 \ucK \eta_2=\mbP\,\eta _1 \uR \eta_2,\eea
   where
   \bea\nn
\mbP&=&\Big[|+\rangle \Big(\langle +|\otimes\langle +|-\langle -|\otimes\langle +|\Big)+|-\rangle\Big(\langle +|\otimes\langle -|+\langle -|\otimes\langle +|\Big )\Big].\eea
Here we used that $\langle -|+\rangle=\langle +|-\rangle=0$
and $\langle +|+\rangle=\langle -|-\rangle=1$.\\

Let us consider the inner product $g$ and the antisymmetric form $\omega$ for two vectors $\eta$ and $\chi$
each of them is the $\cK$ tensor product of $\eta_1, \eta_2$ and $\chi_1, \chi_2$, respectively,
\bea
\eta=\eta_1\ucK \eta_2, \quad  \chi=\chi_1\ucK \chi_2,\eea
or explicitly,
\bea
\eta&=&\eta_1\ucK \eta_2=\sum_{a,b}\begin{pmatrix}
  q_{1a} q_{2b} - p_{1a} p_{2b}\\ q_{1a} p_{2b} +p_{1a} q_{2b}
\end{pmatrix} e_{a}\otimes e_{b}\\
\chi&=&\chi_1\ucK \chi_2=\sum_{a,b}\begin{pmatrix}
  u_{1a} u_{2b} - v_{1a} v_{2b}\\ u_{1a} v_{2b} +v_{1a} u_{2b}
\end{pmatrix} e_{a}\otimes e_{b},
\eea
where $\eta_1,\eta_2$ are given by \eqref{eta12} and
\be
\chi _1=\begin{pmatrix}
   u_1\\v_1  
 \end{pmatrix};\quad 
 \chi _2\begin{pmatrix}
   u_2\\v_2  
 \end{pmatrix}.
\ee

The following relations hold
\bea
\label{g-eta12-K}
g\left(\eta_1\ucK \eta_2,\chi_1\ucK \chi_2\right)&=&g(\eta_1,\chi_1)
g(\eta_2,\chi_2)-\omega(\eta_1,\chi_1)
\omega(\eta_2,\chi_2),\\
\label{o-eta12-K}
\omega\left(\eta_1 \ucK\eta_2,\chi_1\ucK\chi_2\right)
&=&g\left(\eta_1,\chi_1\right)\,\omega\left(\eta_2,\chi_2\right) +
\omega\left(\eta_1,\chi_1\right)\,g\left(\eta_2,\chi_2\right).
\eea

\section{Relations of Hilbert   $\mbC^n$ and   Kähler $\cK^{2n}$ spaces}\label{RHK}

\subsection{From Hilbert to Kähler}\label{HK}
 
To a finite-dimensional complex Hilbert space ${\mathbb C}^n$ with inner product $\langle \cdot,\cdot\rangle $ we associate a real Kähler space  consisting of a real Hilbert space ${\mathbb R}^{2n}$ with inner product $g(\cdot,\cdot)$ which is also equipped with a symplectic form $\omega$ and an automorphism $J$. There is a following relation between the  complex Hilbert space and the real Kähler space:
\begin{itemize}
\item 
we write $\psi=q+ip$, $q\in \mbR^n,$ $p\in \mbR^n$ and $\psi\in \mbC^n$\\
Taking in mind that elements of 
$\cK$ consist of pairs $\vqp$ it is convenient to introduce an operation $\gamma$ that maps pairs of vectors $q\in \mbR^n,$ and $p\in \mbR^n,$ to vector in the form $q+ip\in\mbC^n$ and the operation $\gamma^{-1}$ maping  the complex $\psi=q+ip$ to the pair of its real and imaginary parts
\be
\gamma\vqp=q+ip,\quad \gamma^{-1}(q+ip) =\vqp. 
\ee
Therefore, we have
\be\label{gamma-psi}
\gamma\vqp=\psi, \quad \gamma^{-1}\psi =\vqp.
\ee
\item To specify the structure of the Kähler space on  pairs $\vqp$ we have to define the scalar product and $\omega $ form on these pairs as well as automorphism $J$. We do this using the real and imaginary part of the corresponding scalar product in $\mbC$
\bea\label{g-pairs}
g\left({\vqpf},\vqps\right)&=&\Re \Big(<\gamma \vqps,\gamma\vqps>\Big),\\
\label{o-pairs}
\omega\left({\vqpf},\vqps\right)&=&\Im \Big(<\gamma \vqpf,\gamma\vqps>\Big).\eea
We define
\be \label{JJ}\cJ\vqp=\begin{pmatrix}
    -p\\q
\end{pmatrix}.\ee
\end{itemize}

Since  $\psi_{i} \in \mbC^n=\mbR^n\oplus  i \mbR^n$ has a decomposition $\psi_i = q_i + i p_i$, where $p_i,q_i \in \mbR^n$, $i=1,2$, we can write
\begin{equation}
    \langle \psi_1, \psi_2 \rangle =  \langle q_1 + i p_1,q_2 + i p_2\rangle = \langle q_1,q_2\rangle +\langle p_1,p_2\rangle + i( \langle q_1,p_2 \rangle - \langle p_1,q_2 \rangle).
\end{equation}
and therefore we define $g$ and $\omega$ on the pairs as
\bea
\label{gomega}g\left(\vqpf,\vqps \right) &=& \langle q_1,q_2\rangle +\langle p_1,p_2\rangle,\\
\omega \left(\vqpf,\vqps\right) &=& \langle q_1,p_2 \rangle - \langle p_1,q_2 \rangle.
\label{gomega1}
\eea
One can check explicitly that due to \eqref{gomega}, \eqref{gomega1} and \eqref{JJ}  
the consistency conditions \eqref{g-omega}, \eqref{omega-g} and \eqref{omegagJ}  are fulfilled. 
\\

Therefore, starting from the complex Hilbert space $\mbC^n=\mbR^n\oplus i\mbR^n$ we get the real vector space $\mbR^{2n}=\mbR^n\oplus \mbR^n$ with the complex structure $J$,
\be J:\quad \mbR^{2n} \to \mbR^{2n} \quad\mbox{defined by} \quad\cJ\begin{pmatrix}
    x\\p
\end{pmatrix}=\begin{pmatrix}
    -p\\x
\end{pmatrix},
\ee 
the inner product and the symplectic form, subject of consistency relations, i.e. we construct the real Kähler space $(\mbR^n\oplus \mbR^n, g,\,\omega,\, J)$. It is obsious that  in the ortogonal basis we get 
\eqref{com-g},\eqref{com-omega} and  \eqref{mCS}.
\\

Since an arbitrary Hermitian matrix in $\mbC$ has a decomposition on real symmetric matrix and real antisymmetric one, given by the formula \eqref{LSA},
due to the relation
\be
L\psi=(S+i A)(q+ip)=(S q- Ap)+i(A q+Sp)\ee
we get the corresponding operator $\cL$ acts on pairs as the block matrix
\be \cL\vqp=\begin{pmatrix}
    S&-A\\A&S
\end{pmatrix}\,\vqp\ee

\subsection{From Kähler to Hilbert}

Starting from the Kähler space \((\mathcal{K}, g, \omega, J)\), we construct a complex Hilbert space \((\mathcal{H}, \langle \cdot, \cdot \rangle)\).  
Let \(\mathcal{K} = K \oplus K\), and define elements \(\psi \in \mathcal{H}\) as \(\psi = q + i p\), where \(q, p \in K\). The inner product on \(\mathcal{H}\) is given by
\begin{equation}
    \langle \psi_1, \psi_2 \rangle = g\big(\gamma^{-1}\psi_1, \gamma^{-1}\psi_2\big) + i \, \omega\big(\gamma^{-1}\psi_1, \gamma^{-1}\psi_2\big),
\end{equation}
where \(g\) and \(\omega\) are the metric and symplectic form defined in \eqref{gomega} and \eqref{gomega1}, respectively.  
This construction yields a complex Hilbert space, which corresponds to the complex structure on \(\mathcal{K}\).

\subsection{Comparison of tensor product $\uC $ on Hilbert space $\mbC^{n}$ and 
 \\ tensor product $\ucK $ on the Kahler space $\cK^{2n}$}

  In $\mbC ^{n}$ there is a basis $\{e_{a}, a=1,...n\}$
   and the tensor product of two elements $\psi=\sum_a (q_{1a} +
 ip_{1a}) e_{a}$ and $\phi=\sum_b \left(q_{2b}+i 
 p_{2b}\right) e_{b}$ is 
   \bea
\label{1ten}
\psi \uC \phi=\sum_{a,b} \Big(q_{1a} q_{2b} - p_{1a} p_{2b}+ i(q_{1a} p_{2b} +p_{1a} q_{2b})\Big) e_{a}\otimes e_{b}. \eea

Let pairs of real $\eta$ and $\chi$
correspond to the complex wave functions
$\psi_{\eta}$ and $\psi_{\chi}$
\bea
\psi_{\eta_i}=\gamma \,\eta_i\quad
\psi_{\chi_i}=\gamma \,\chi_i,\eea
where $\gamma$ is defined as in \eqref{gamma-psi}. 
We can say that $\psi_\eta$
and $\psi_\chi$ compose on two subsystem $\psi_{\eta_i}, i=1,2$
and $\psi_{\chi,i},i=1,2$. respectively, i.e.
\bea
\psi_{\eta}&=&\psi_{\eta_1}\psi_{\eta_2},\\
\psi_{\chi}&=&\psi_{\chi_1}\psi_{\chi_2},
\eea
and we have
\bea
\label{psi1psi2}\langle\psi_\eta,\psi_\chi\rangle&=&
\langle\psi_{\eta_1}\psi_{\eta_2},\psi_{\chi_1} \psi_{\chi_2}\rangle=\langle\psi_{\eta_1},\psi_{\chi_1}\rangle \,\langle\psi_{\eta_2},\psi_{\chi_2} \rangle.
\eea
Taking into account \eqref{g-pairs} and \eqref{o-pairs}
we rewrite the RHS of \eqref{psi1psi2} as 
\bea\nn
\langle\psi_{\eta_1},\psi_{\chi_1}\rangle \,\langle\psi_{\eta_2},\psi_{\chi_2} \rangle
&=&\Big(g(\eta_1,\chi_1)+i\omega(\eta_1,\chi_1)\Big)
\Big(g(\eta_2,\chi_2)+i\omega(\eta_2,\chi_2)\Big)\\
&=&g(\eta_1,\chi_1)g(\eta_2,\chi_2)-\omega(\eta_1,\chi_1)
\omega(\eta_2,\chi_2)\nn\\&+&i
\Big(g(\eta_1,\chi_1)\omega(\eta_2,\chi_2)+\omega(\eta_1,\chi_1)
g(\eta_2,\chi_2)\Big).\eea
Applying relations \eqref{g-pairs} and \eqref{o-pairs} 
to the LHS of \eqref{psi1psi2}
we get
\bea\label{g-com}
g\left(\eta_1\uR \eta_2,\chi_1\uR \chi_2\right)&=&g(\eta_1,\chi_1)g(\eta_2,\chi_2)-\omega(\eta_1,\chi_1)
\omega(\eta_2,\chi_2)\nn\\
\label{omega-com}\omega \left(\eta_1\uR \eta_2,\chi_1\uR \chi_2\right)&=&
(g(\eta_1,\chi_1)\omega(\eta_2,\chi_2)+\omega(\eta_1,\chi_1)
g(\eta_2,\chi_2),\eea
that are nothing but relations \eqref{g-eta12-K}
and \eqref{o-eta12-K}.

\section{Quantum Mechanics in Real Kähler Space}
\label{QMK}

 In this Section we  formulate the main principles of quantum mechanics directly in the real Kähler space. 
 In real Kähler space formulation, quantum theory is defined in terms of the following postulates~\cite{Volovich:2025rmi}:
 
\begin{itemize}
 \item i) To a physical system one assigns a real Kähler space $\cK$ (instead of the a Hilbert space $\cH$)
 and its state is represented by  normalized  vectors  $\eta\in \cK$, that is, $g(\eta,\eta)=1$, where $g(\cdot,\cdot)$ is an inner product in $\cK$. 
\item  ii) To the observable $L$ corresponds  the $\cK$
Hermitian operator $\cL$, which spectrum is observable.  The spectral decomposition for $\cL$ is given by \eqref{EL}, \eqref{EI}:
$
  \sum_{i=1}^{n} \lambda_i \cE_{i}= \cL$, $ \sum_{i=1}^{n} \cE_i = \cI$ and $\mbox{rank} (\cE_i)\geq 2$.
\item iii) Born rule: if we measure $L$   in the normalized state $\eta$, the probability of obtaining result $\lambda_i$ is given by $g(\eta, \cE_i\eta)/\mbox{rank} (\cE_i)$.
 
\item iv) the Kahler space $\cK$ corresponding to the composition of two systems $\fN$ and $ \fM$ is $\cK_\fN\ucK\cK_\fM$.

\item v$^*$)\footnote{Postulates marked with * are optional.} A compact Lie group $\fG$ of internal symmetries is realized in the Kähler space $\cK$ by a symplectic orthogonal representation $\cU(g)$ for $g \in \fG$.

\item vi$^*$) The Galilei or Poincaré group of space-time transformations  $\fP$ is realized in the Kähler space $\cK$ by a symplectic orthogonal representation $\cU(p)$ for $p \in \fP$.
\end{itemize}

Note also   the consideration of quantum states as equivalence classes under phase transformations $\cU(1)$. Dealing with the Kähler space
maintaining phase invariance through equivalence class definitions of states.

\section{Comparison of Quantum Mechanics in $\mathbb{C}^n$ and $\mathcal{K}^{2n}$ Spaces}\label{QMKH}
There is a finite number of parties, each described by a Hilbert space $\mathbb{C}^m$. These parties form direct sums and tensor products, transmit quantum messages to each other, and perform quantum measurements and computations. Ultimately, an entire network (scheme) reduces to the study of quantum mechanics in the space $\mathbb{C}^n$ with $m \leq n$.
The scheme may encompass such areas of quantum information as quantum communication, quantum cryptography, quantum computing, and quantum nonlocality. Generally speaking, all results of experiments can be read off \cite{Dirac,Landau:1991wop} from the following expression in the complex Hilbert space $\mathbb{C}^n$:
\begin{equation}\label{5.1}
\langle L_1 \cdots L_k \psi, \phi \rangle.
\end{equation}

In \cite{Volovich:2025rmi} it was shown that expression \eqref{5.1} can be obtained from data encoded in the real Kähler space. For this purpose, a natural isomorphism $\gamma$ between operator theories in complex Hilbert space and real Kähler space is defined \cite{Volovich:2025rmi}, yielding the relation:
\begin{align}\label{5.2}
\langle L_1 \cdots L_k \psi, \phi \rangle &= 
g\bigl(\mathcal{L}_1 \cdots \mathcal{L}_k(q,p),(r,s)\bigr) + i \omega\bigl(\mathcal{L}_1 \cdots \mathcal{L}_k(q,p),(r,s)\bigr), \\ \label{5.3}
\gamma^{-1}\psi &= \vqp, \quad \gamma^{-1}\phi = \rqs,
\end{align}
where $L_i$ are operators in the Hilbert space and $\mathcal{L}_i$ are operators in the Kähler space (see Sect.~\ref{RHK}). In particular, we associate natural partners in real Kähler space to each Hermitian, unitary, and projection operator in complex Hilbert space.

This relation shows that all principal results obtained in complex Hilbert space quantum mechanics can be reproduced using corresponding 
              real Kähler space quantum mechanics. 
 This result will be called the reconstruction theorem.
\\

\noindent\textbf{Theorem.} In the Hilbert space $\mathbb{C}^n$ with the standard inner product $\langle \cdot, \cdot \rangle$, consider the correlation function \eqref{5.1} where $L_i$ are Hermitian, unitary, or projection operators. It can be represented in the form \eqref{5.2}, where $g(\cdot,\cdot)$ and $\omega(\cdot,\cdot)$
are given by \eqref{g-pairs} and \eqref{o-pairs}
with relations of $\psi$ and $\phi$ with $(q,p)$ and (r,s) as in \eqref{5.3}.

 \newpage.
 \section{Conclusion and Discussion}

In this work, we have demonstrated the equivalence between the complex Hilbert space formulation of quantum mechanics and its real Kähler space counterpart.
Our Kähler-space framework resolves ambiguities in  debates on real vs. complex formulations by:
\begin{enumerate}
    \item Avoiding ad hoc assumptions: Composite systems are treated via standard tensor products in $\mbR^{2n}$, without introducing hidden variables or nonlocal maps, but by introducting the projection operator that makes one to one correspondence between wave function of composite system in the Hilbert and Kähler spaces. 
    
    \item Preserving spectral equivalence: Operator degeneracy in $\cK^{2n}$ mirrors the algebraic structure of Hermitian operators in $\mbC^n$.
\end{enumerate}

 While debates on the "reality" of complex numbers persist, our results demonstrate that real Kähler spaces provide a consistent, self-contained framework for real quantum mechanics. Future directions include:

\begin{itemize}
    \item Experimental tests of real-complex discrepancies in network scenarios.
    \item Computational advantages of the Kähler formulation.
    \end{itemize}

\section*{Acknowledgment}
 I are grateful to   R. Singh,  A.~Teretenkov and A.~Trushechkin for useful discussions.
This work is supported by the Russian Science Foundation (24-11-00039, Steklov Mathematical Institute).

\newpage
\appendix
\section{Proof of algebraic relations concerning the tensor\\ products}
\subsection{Prof of formula for scalar product and symplecting form for vectors that are    tensor $\uR$ product in $\cK^{2n}$}
Let's prove  the formula  \eqref{TRg} and  \eqref{TRo}.
Using representation  \eqref{T12R} we get
 \bea
  &&\qquad\qquad g\left( \eta _1 \uR \eta_2, \chi _1 \uR \chi_2\right)\\\nn&=&
\sum_{a,b} \Big(q_{1a} q_{2b}
 u_{1a} u_{2b} +p_{1a} q_{2b} v_{1a} u_{2b}+q_{1a} p_{2b}
 u_{1a} v_{2b} +
 p_{1a} p_{2b} v_{1a} v_{2b}\Big)\\
  \nn
&=& \Big(\sum_{a} (q_{1a} u_{1a}+p_{1a} v_{1a}) \Big)
\Big(\sum_{b} ( q_{2b}
 u_{2b}+ p_{2b}
  v_{2b} ) \Big )=g\left( \eta _1 ,\chi _1 )\cdot g(\eta_2, \chi_2\right),\eea
  here we used that $\langle -|+\rangle=\langle +|-\rangle=0$
and $\langle +|+\rangle=\langle -|-\rangle=1$.
\\

We also have
\bea \omega\left( \eta _1 \uR \eta_2, \chi _1 \uR \chi_2\right)=
\omega\left( \eta _1 ,\chi _1 )\cdot g(\eta_2, \chi_2\right)+g\left( \eta _1 ,\chi _1 )\cdot \omega(\eta_2, \chi_2\right).
\eea

\subsection{Prof of formula for scalar product and symplecting form for vectors that are    tensor $\ucK$ product in $\cK^{2n}$}
Let us prove the properties
\eqref{g-eta12-K} and \eqref{o-eta12-K}. We take
\bea
\eta=\eta_1\ucK \eta_2, \quad  \chi=\chi_1\ucK \chi_2,\eea
or explicitly,
\bea
\eta&=&\eta_1\ucK \eta_2=\sum_{a,b}\begin{pmatrix}
  q_{1a} q_{2b} - p_{1a} p_{2b}\\ q_{1a} p_{2b} +p_{1a} q_{2b}
\end{pmatrix} e_{a}\otimes e_{b}\\
\chi&=&\chi_1\ucK \chi_2=\sum_{a,b}\begin{pmatrix}
  u_{1a} u_{2b} - v_{1a} v_{2b},\\ u_{1a} v_{2b} +v_{1a} u_{2b}
\end{pmatrix} e_{a}\otimes e_{b},
\eea
where $\eta_1,\eta_2$ are given by \eqref{eta12} and
\be
\chi _1=\begin{pmatrix}
   u_1\\v_1  
 \end{pmatrix};\quad 
 \chi _2\begin{pmatrix}
   u_2\\v_2  
 \end{pmatrix}.
\ee 

We have
\bea
\label{T12CRT34}
&&g\left(\eta_1\ucK \eta_2,\chi_1\ucK \chi_2\right)\\\nn
&=&g\Big( \sum_{a,b}
 \Big( (q_{1a} q_{2b} - p_{1a} p_{2b})|+\rangle+( q_{1a} p_{2b} +p_{1a} q_{2b})|-\rangle\Big)\otimes
 e_{a}\otimes e_{b},\\\nn
 &&\sum_{a,b}
 \Big( (u_{1a} u_{2b} - v_{1a}v_{2b})|+\rangle+( u_{1a} v_{2b} +v_{1a} u_{2b})|-\rangle\Big)\otimes
 e_{a}\otimes e_{b}\Big)
\\ \nn&=& \sum_{a,b}
 \Big( (q_{1a} q_{2b} - p_{1a} p_{2b}) 
 (u_{1a} u_{2b} - v_{1a} v_{2b})\Big)+
 \sum_{a,b}
 \Big(( q_{1a} p_{2b} +p_{1a} q_{2b})( u_{1a} v_{2b} +v_{1a} u_{2b})\Big)
\\\nn&=&
 \sum_{a}\Big( q_{1a}u_{1a}+p_{1a}v_{1a} \Big) \sum_{b}\Big( q_{2b} u_{2b}+p_{2b} v_{2b}\Big)+\sum_{a}\Big(q_{1a} v_{1a}-  p_{1a}u_{1a}\Big) \sum_{b}\Big(p_{2b} u_{2b}-q_{2b} v_{2b}\Big)\\
\nn&=&g\left(\begin{pmatrix}
   q_1\\u_1  
\end{pmatrix},\begin{pmatrix}
   u_1\\v_1  
\end{pmatrix}\right)\cdot g\left(\begin{pmatrix}
   q_2\\u_2  
 \end{pmatrix},\begin{pmatrix}
   u_2\\v_2  
 \end{pmatrix}\right)-
 \omega\left(\begin{pmatrix}
   q_1\\u_1  
 \end{pmatrix},\begin{pmatrix}
   u_1\\v_1  
\end{pmatrix}\right)\cdot \omega\left(\begin{pmatrix}
   q_2\\u_2  
 \end{pmatrix},\begin{pmatrix}
   u_2\\v_2  
 \end{pmatrix}\right),
\eea
that can be   rewrite  as
\be
g\left(\eta_1\ucK \eta_2,\chi_1\ucK \chi_2\right)=g(\eta_1,\chi_1)
g(\eta_2,\chi_2)+\omega(\eta_1,\chi_1)
\omega(\eta_2,\chi_2).\ee
We also have
\bea
\label{T12CRT34omega}
&&\omega \left(\eta_1\ucK \eta_2,\eta_1\ucK \eta_2\right)\\\nn
&=&\omega\Big( \sum_{a,b}
 \Big( (q_{1a} q_{2b} - p_{1a} p_{2b})|+\rangle+( q_{1a} p_{2b} +p_{1a} q_{2b})|-\rangle\Big)\otimes
 e_{a}\otimes e_{b},\\\nn
 &&\sum_{a,b}
 \Big( (u_{1a} u_{2b} - v_{1a}v_{2b})|+\rangle+( u_{1a} v_{2b} +v_{1a} u_{2b})|-\rangle\Big)\otimes
 e_{a}\otimes e_{b}\Big)
 \\&=&
 \sum_{a,b}
 \Big(\Big( (q_{1a} u_{1a} +p_{1a} v_{1a})(q_{2b}v_{2b}-u_{2b} p_{2b})+ (q_{2b}u_{2b}+p_{2b}v_{2b})( q_{1a}
 v_{1a}-u_{1a}p_{1a} )\nn\\
 \nn&=&g\left(\begin{pmatrix}
   q_1\\u_1  
 \end{pmatrix},\begin{pmatrix}
   u_1\\v_1  
\end{pmatrix}\right)\cdot \omega\left(\begin{pmatrix}
   q_2\\u_2  
 \end{pmatrix},\begin{pmatrix}
   u_2\\v_2  
 \end{pmatrix}\right)+ g\left(\begin{pmatrix}
   q_2\\u_2  
 \end{pmatrix},\begin{pmatrix}
   u_2\\v_2  
 \end{pmatrix}\right)
 \omega\left(\begin{pmatrix}
   q_1\\u_1  
 \end{pmatrix},\begin{pmatrix}
   u_1\\v_1  
\end{pmatrix}\right),\eea
\\
that can be written as 
\bea
\omega\left(\eta_1 \ucK\eta_2,\chi_1\ucK\chi_2\right)
&=&g\left(\eta_1,\chi_1\right)\,\omega\left(\eta_2,\chi_2\right) +
\omega\left(\eta_1,\chi_1\right)\,g\left(\eta_2,\chi_2\right).
\eea

\section{Examples of $\mbC^n$ and $\cK ^{2n}$ correspondence}\label{Cn-K2n}

\subsection{$\mbC^1$ and $\cK^2$} We start with the real $2\times 2$ symmetric matrices
\be
\ell=\left(
\begin{array}{cc}
 s_1 & a  \\
 a & s_2 
\end{array}
\right).\ee
The complex structure matrix is 
\be
j=\left(
\begin{array}{cc}
 0 & -1  \\
 1 & 0 
\end{array}
\right)\ee
and the second  condition  in \eqref{cLJ}, $\ell j-j\ell=0$, implies that $a=0$, $s_1=s_2\equiv s$ and we left with the diagonal matrix in the form \be \ell_0=\left(
\begin{array}{cc}
 s & 0  \\
 0 & s 
\end{array}
\right)\ee with equal eigenvalues 
\be
\lambda _{1,2}=s.\ee
We have two projectors on corresponding eigenvectors
\be
P_1=\left(
\begin{array}{cc}
 1 & 0  \\
 0 & 0 
\end{array}
\right),\quad P_2=\left(
\begin{array}{cc}
 0 & 0  \\
 0 & 1 
\end{array}
\right),\ee
that satisfy 
\be jP_1- P_2j=0.\ee
The resolution of the identity has a simple form 
\be 
P_1+P_2=1,\quad s(P_1+P_2)=\ell_0.\ee

 \subsection{$\mbC^2$ and $\cK^4$}\label{C2-Bloch}
 As an example of the relation of Hilbert and Kähler spaces let us consider the model in the Hilbert space $\mbC^2$ which as  related with a model in $\cK^4$.
 \subsubsection{$\mbC^2$ and  a Qubit}
 Let us remind that a 2-dimensional complex Hilbert space $\mbC^2$ is considered as  a qubit (a two-level quantum system). The  basis states in $\mbC^2$ are  \footnote{To distinguish these basis vectors from the 2-component basis in $\mbR$ we use notations $|0\rangle,|1\rangle$ instead of 
   $|+\rangle,|-\rangle$ in $\mbR^2$}
$$
|0\rangle = \begin{pmatrix} 1 \\ 0 \end{pmatrix}, \quad |1\rangle = \begin{pmatrix} 0 \\ 1 \end{pmatrix}.$$
 
A pure state is a superposition:
\[
|\psi\rangle = \alpha|0\rangle + \beta|1\rangle \quad \text{with } \alpha, \beta \in \mathbb{C} \text{ and } |\alpha|^2 + |\beta|^2 = 1.
\]
Ignoring global phase (\( |\psi\rangle \sim e^{i\theta}|\psi\rangle \)), the state space becomes the Bloch sphere \( S^2 \) with parametrization:
\be\label{Psi}
|\psi\rangle = \cos\left(\frac{\theta}{2}\right)|0\rangle + e^{i\phi}\sin\left(\frac{\theta}{2}\right)|1\rangle,
\ee
where \( \theta \in [0, \pi] \) and \( \phi \in [0, 2\pi) \).

A mixed state is described by a density matrix \( \rho \), which is positive semi-definite, Hermitian, and satisfies \( \text{Tr}(\rho) = 1 \). For a qubit:
\[
\rho = \frac{1}{2}\left(I + r_x\sigma_x + r_y\sigma_y + r_z\sigma_z\right),
\]
where \( \mathbf{r} = (r_x, r_y, r_z) \) is the Bloch vector with $ ||\mathbf{r}|| \leq 1 $. The set of all states forms the Bloch ball \( B^3 \):
$||\mathbf{r}|| \leq 1$. Therefore, 
a pure states form the surface ($ ||\mathbf{r}||= 1 $) the mixed states form the interior ($ ||\mathbf{r}|| < 1 $).
A pure qubit state has two degrees of freedom: the angles 
$\varphi$   and 
$\theta $.
A mixed qubit state has three degrees of freedom: the angles 
$\varphi$   and 
$\theta $, as well as the length 
$r$ of the vector that represents the mixed state.
\\

The Bloch sphere \( S^2 \), representing pure states of a qubit, can indeed be endowed with a complex structure. This arises from its identification with the  complex projective line \( \mbC P^1 \), a 1-dimensional complex manifold. Or in other words the Bloch sphere admits the Kahler structure.
\\

The generalization  the Bloch sphere to  $n>2$ is considered in \cite{Ashtekar:1997ud}, see also \cite{Kible}.

 \subsubsection{Eigenvalues problem in $\mbC^2$ }
Let us consider the following general Hermitian matrix acting in $\mbC^2$ 
\be\label{L2}
L=\left(
\begin{array}{cc}
 s_{11} & s_{12}+i
   a \\
 s_{12}-i a &
   s_{22} \\
\end{array}
\right)\ee
Here $s_{ik},\,i=1,2$ and $a$ are real.
It can be representing as 
\be\label{LSA}
L=S+iA,\quad \mbox{where}\quad 
S=\left(
\begin{array}{cc}
 s_{11} & s_{12}
    \\
 s_{12}&
   s_{22} \\
\end{array}
\right),\quad A=\left(
\begin{array}{cc}
 0 & 
   a \\
 -a &
   0\\
\end{array}
\right)\ee

The matrix $L$ has the following orthonormal eigenvectors $V_{\alpha},\, \alpha=1,3$
\bea
V_1&=&n_-
\left(
\begin{array}{c}
 +!\frac{w_-(w_0+i)}
  {1+w_0^2
   }\\
 1 \\
\end{array}
\right), \quad (V_1,V_1)=1
\\V_2&=&n_+\left(
\begin{array}{c}
 \frac{w_+
   \left(w_0+i\right)}{1+w_0^2
   } \\
 1 \\
\end{array}
\right), \quad (V_2,V_2)=1\quad (V_1,V_3)=0,
\eea
and their corresponding eigenvalues are given by \eqref{lambda1} and \eqref{lambda3} 
\bea
L\,V_1 &=&\lambda _1 V_1,\\
L\,V_2 &=&\lambda _2 V_2\eea
Here \bea 
  \label{nw}
w_{\pm}&=&\frac{\pm\kappa+s_{11}-s_{22}}{2 a}, \quad 
w_0= \frac{s_{12}}{a},\quad n_\pm=\frac{\sqrt{1+w_0^2}}{\sqrt{1+w_\pm^2+w_0^2}},\\
\label{kappa}\kappa&=&\sqrt{4 a^2+s_{11}^2-2 s_{11}
   s_{22}+4 s_{12}^2+s_{22}^2},
\eea
and eigenvalues are
\bea\label{lambda1}
\lambda_1&=&\frac{1}{2} (-\kappa+s_{11}+s_{22}),\\\label{lambda2}
\lambda_2&=&\frac{1}{2} (\kappa+s_{11}+s_{22}).\eea

Using representation \eqref{LSA} the action of $L$ on a two-component complex vector $Z$
\be\label{Z}
Z=\begin{pmatrix}
  z_1\\ z_2 
\end{pmatrix}=\begin{pmatrix}
  x_1+ip_1\\ x_2 +ip_2
\end{pmatrix}=\begin{pmatrix}
  x_1\\ x_2 
\end{pmatrix}+i\begin{pmatrix}
  p_1\\p_2
\end{pmatrix}=X+i P\ee
  can be written as
\be
\label{Z'}
Z'=LZ=X'+i P'=(S+iA)(X+i P)=
S X- A P +i (A X+ S P)
\ee
and we have
\be \label{X'Y'}X'=S X- A P,\quad P'=A X+ S P.\ee
Or in two-component notations
\be \begin{pmatrix}
    X'\\P'
\end{pmatrix}=\begin{pmatrix}
    S&-A\\A&S
\end{pmatrix}\,\begin{pmatrix}
    X\\P
\end{pmatrix}\ee

 \subsubsection{Eigenvalues problem in $\mbR^4$ }
 The general form of $4\times 4$ real symmetric matrices commuting with the matrix $\cJ$ representing the complex structure, i.e. the matrix satisfying conditions  of the form \eqref{cL}, is
 \be\label{cL4}
\cL_4=\left(
\begin{array}{cccc}
 s_{11} & s_{12} & 0 & -a \\
 s_{12} & s_{22} & a & 0 \\
 0 & a & s_{11} & s_{12} \\
 -a & 0 & s_{12} & s_{22} \\
\end{array}
\right).\ee
Eigenvalues of the matrix $\cL_4$ are 
\bea\label{lambda1m}
\lambda_1&=&\lambda_3\\\label{lambda3m}
\lambda_2&=&\lambda_4\eea
where $\lambda_1$ and $\lambda_2$ are given by \eqref{lambda1} and \eqref{lambda2}.
   
We denote four eigenvectors as 
\bea
\cV_i&=&\left(
\begin{array}{c}
 v_{i1} \\
 v_{i2} \\
 v_{i3}\\
 v_{i4}
\end{array}
\right),\quad i=1,2,3,4.
\eea
The two first orthogonal eigenvectors corresponding to 
eigenvalues  $\lambda_1$ and $\lambda_2$ are 
\bea 
\cV_1&=&\gamma (V_1)=n_-\gamma 
\left(
\begin{array}{c}
 \frac{w_-(w_0+i)}
  {1+w_0^2
   }\\
 1 \\
\end{array}
\right)=n_-
\left(
\begin{array}{c}
   w_0\rho_-\\
 1 \\
 \rho_- \\
 0 \\
\end{array}
\right),\\\cV_{2}&=&\gamma(V_2)=n_+ \gamma\left(
\begin{array}{c}
 \frac{w_+
   \left(w_0+i\right)}{1+w_0^2
   } \\
 1 \\
\end{array}
\right) =n_+\left(
\begin{array}{c}
w_0\rho_+ \\
 1 \\
 \rho_+ \\
 0 \\
\end{array}
\right),\eea 
where
\bea
 \rho_\pm=\frac{
   w_\pm}{
(1+w_0^2) }
\eea
and $w_{\pm}, w_0, n_\pm$ are given by \eqref{nw} and 
\eqref{kappa}.

$\cV_{3}$ and $\cV_{4}$ can be obtained by acting of the complex operator $\cJ$ on $\cV_{1}$ and $\cV_{2}$, respectively,
\bea \cV_{3}=\cJ\,\cV_{1},\quad
\cV_{4}=\cJ\,\cV_{2},
\eea 
Therefore, the spectrum is  double degenerated 
\bea
(\cL -\lambda_1)\cV_1&=&0\,\quad
(\cL -\lambda_2)\cV_2=0\\
 (\cL -\lambda_1)\cV_3&=&0,\quad
(\cL -\lambda_2)\cV_4=0.\eea
This property follows immediately from 
\eqref{cLJ}.

These vectors $\cV_i$, $i=1,...4$ are orthonormal 
\bea
\cV_1^{\rT} \cdot \cV_1&=&\cV_2^{\rT} \cdot \cV_2=\cV_3^{\rT} \cdot \cV_3=\cV_4^{\rT} \cdot \cV_4=1;\\
\cV_1^{\rT} \cdot \cV_2&=&\cV_1^{\rT} \cdot \cV_3=\cV_1^{\rT} \cdot \cV_4=0;\\
\cV_2^{\rT} \cdot \cV_3&=&\cV_2^{\rT} \cdot \cV_4=0;\\
\cV_3^{\rT} \cdot \cV_4&=&0.
\eea
here 
\be \cV_i^{\rT}=(v_{i1},v_{i2},v_{i3},v_{i4}), \qquad \cV_i^{\rT}\cdot \cV_j=\sum _{m=1}^4v_{im}v_{jm}.\ee

Defining 
\bea
\cP_1&=&\cV_1 \otimes \cV_1^{\rT};\qquad \cP_3=-\cJ\cP_1\cJ\\
\cP_2&=&\cV_2 \otimes \cV_2^{\rT};\qquad \cP_4=-\cJ\cP_2\cJ\eea
we get the following spectral decomposition 
\bea
\sum _{i=1}^{4}\lambda_i \cP_i =\cL_4, \qquad \sum _{i=1}^{4}\cP_i = \cI_4\eea
here $\cI_4$ is the unit $4\times 4$ matrix.
\\

\subsubsection{Representations
\eqref{cL4} and \eqref{L2}}
Note the connection of the representations
\eqref{cL4} and \eqref{L2} follows from the following consideration. According our scheme described in Sect.\ref{Sect:intro}, see also \cite{Volovich:2025rmi}, to the vector $Z$
in $\mbC^2$ given in \eqref{Z}
corresponds the vector in 
$\mbR^4$
\bea\label{Z24}
Z=\begin{pmatrix}
  x_1+ip_1\\ x_2 +ip_2
\end{pmatrix}
\Leftrightarrow
\cZ=\begin{pmatrix}
  x_1\\ x_2 \\p_1\\p_2
\end{pmatrix}\eea
One can check explicitly that the transformation \eqref{Z'}
\bea
\label{Z'X'P'}
LZ=Z'&=&\begin{pmatrix}
  x_1'+ip_1'\\ x_2' +ip_2'
\end{pmatrix},\eea where 
\bea \label{x1'}
x_1'&=&s_{11} x_1+s_{12} x_2- a_{12} p_2,\\\label{x2'}
x_2'&=&s_{21} x_1+s_{22} x_2- a_{21} p_1,\\
\label{p1'}p_1'&=& a_{12} x_2+ s_{11} p_1+ s_{12} p_2,
\\
\label{p2'}p_2'&=& A_{21} x_1+ S_{21} p_1+ S_{22} p_2,
\eea
corresponds to 
\bea
\cZ'=\begin{pmatrix}
  x_1'\\ x_2' \\p_1'\\p_2'
\end{pmatrix}=\left(
\begin{array}{cccc}
 s_{11} & s_{12} & 0 & -a \\
 s_{12} & s_{22} & a & 0 \\
 0 & a & s_{11} & s_{12} \\
 -a & 0 & s_{12} & s_{22} \\
\end{array}
\right)\begin{pmatrix}
  x_1\\ x_2 \\p_1\\p_2
\end{pmatrix},\eea
where $a_{12}=-a$ and  $a_{21}=a$.

\subsubsection{Relations between $U(2)$ rotations in $\mbC^2$ and $O(4)\cap Sp(4)$ transformations in $\cK^4$}.
\bea
\begin{aligned}
&G_1 = \begin{pmatrix}
0 & 1 & 0 & 0 \\
-1 & 0 & 0 & 0 \\
0 & 0 & 0 & 1 \\
0 & 0 & -1 & 0
\end{pmatrix}, \quad
G_2 = \begin{pmatrix}
0 & 0 & -1 & 0 \\
0 & 0 & 0 & -1 \\
1 & 0 & 0 & 0 \\
0 & 1 & 0 & 0
\end{pmatrix}, \\
&G_3 = \begin{pmatrix}
0 & 0 & 0 & -1 \\
0 & 0 & 1 & 0 \\
0 & -1 & 0 & 0 \\
1 & 0 & 0 & 0
\end{pmatrix}, \quad
G_4 = \begin{pmatrix}
0 & -1 & 0 & 0 \\
1 & 0 & 0 & 1 \\
0 & 0 & 0 & 1 \\
0 & 0 & -1 & 0
\end{pmatrix}.
\end{aligned}\eea
These generators span the Lie algebra of 
$ O(4)\cap Sp(4)$.

 These generators possess {the following properties:

\begin{itemize}
    \item Skew-Symmetry:  
    All generators satisfy \( G_i^T = -G_i \), ensuring their membership in \( \mathfrak{o}(4) \).

    \item Preservation of the Symplectic Form:  
    For each \( G_i \), the condition \( G_i J = J G_i \) holds, where  
    \[
    \cJ = \begin{pmatrix} 0 & I \\ -I & 0 \end{pmatrix},
    \]  
    guaranteeing membership in \( \mathfrak{sp}(4) \).

    \item Isomorphism with \( U(2) \):  
    \begin{itemize}
        \item Generators \( G_1, G_2, G_3 \) correspond to \( SU(2) \).
        \item Generator \( G_4 \) corresponds to \( U(1) \), note that $G_2=-\cJ$ .
    \end{itemize}
\end{itemize}
The group itself consists of matrices formed by exponentiating linear combinations of these generators:
\be
M=\exp\{\sum_{i=1}^4\theta _i G_i\},\ee
where $\theta_i\in \mbR$. This structure mirrors the unitary group 
$U(2)$ embedded in 
$\mbR^{4\times 4?}$

 We consider the transformation 
 \be
 Z->Z'=e^{i \phi }Z,\ee
 or in components
 \bea\label{xp'}
&&\qquad\qquad\left(
\begin{array}{c}
 x_1'+i p_1' \\
 x_2'+i p_2' \\
\end{array}
\right)=e^{i \phi}\left(
\begin{array}{c}
 x_1+i p_1 \\
 x_2+i p_2 \\
\end{array}
\right)\\\nn&=& \left(
\begin{array}{c}
x_1 \cos
   (\varphi ) -p_1 \sin (\varphi )+i p_1
   \cos (\varphi )+i x_1 \sin
   (\varphi ) \\
x_2 \cos
   (\varphi )  -p_2 \sin (\varphi )+i p_2
   \cos (\varphi )+i x_2 \sin
   (\varphi )\\
\end{array}
\right),\eea
and 
\bea
\label{z'}
x_1 '&=&x_1 \cos (\varphi )-p_1 \sin
   (\varphi ),\\
   x_2 '&=&x_2 \cos (\varphi )-p_2 \sin
   (\varphi ),\\
   p_1 '&=&p_1 \cos (\varphi )+x_1 \sin
   (\varphi ),\\
   p_2 '&=&p_2 \cos (\varphi )+x_2 \sin
   (\varphi ),
  \eea
   that can be represented as the action of the matrix $g_2(\varphi)$ given by 
    \be g_2(\varphi)=\exp\{\varphi\, G_2 \}.
 \ee
 on 4-component vector $\cZ$ given in \eqref{Z24}, i.e. $\cZ'=g_2(\varphi)\cZ$ Indeed,
\be g_2(\varphi)=
\left(
\begin{array}{cccc}
 \cos (\varphi ) & 0 & -\sin
   (\varphi ) & 0 \\
 0 & \cos (\varphi ) & 0 &
   -\sin (\varphi ) \\
 \sin (\varphi ) & 0 & \cos
   (\varphi ) & 0 \\
 0 & \sin (\varphi ) & 0 &
   \cos (\varphi ) \\
\end{array}
\right)
\ee
and
 \bea
   \begin{pmatrix}
    x_1 '\\x_2 '\\p_1 '\\p_1 '
\end{pmatrix}=g_2(\varphi)\begin{pmatrix}
    x_1 \\x_2 \\p_1 \\p_1 
   \end{pmatrix}.
   \eea

\subsubsection{Connections of different systems of eigenvectors  for
\eqref{cL4} and \eqref{L2}}
Mathematica gives the different set of eigenvectors, we denote them $\cU_i$, $i=1,2,3,4$
\bea\label{B50}
\cU_1=\begin{pmatrix}
  -w_{-}\\
   -w_0\\
   0\\1
\end{pmatrix},\quad \cU _2=\begin{pmatrix}
    w_0\\-w_+\\1\\0
\end{pmatrix},\quad
\cU _3=\begin{pmatrix}
    -w_+\\-w_0\\0\\1
\end{pmatrix}, \quad \cU _4=\begin{pmatrix}
    w_0\\-w_-\\1\\0
\end{pmatrix}
\eea
and
\bea
\cL\,\cU _1-\lambda _1\cU _1=0, \quad
\cL\,\cU _2-\lambda _1\cU _2=0,\\
\cL\,\cU _3-\lambda _2\cU _3=0,\quad
\cL\,\cU _4-\lambda _2\cU _4=0.\eea
Note that not all these vectors are orthogonal
\bea
(\cU _1,\cU _2)&=&(w_+ -w_- )w_0=
\frac{\kappa s_{12}
   }{a^
   2},
\\(\cU _3,\cU _4)&=&(w_- -w_+) w_0=-\frac{\kappa
   s_{12}}{a^2},\\
(\cU _1,\cU _3)&=&(\cU _1,\cU _4)=(\cU _3,\cU _2)=(\cU _4,\cU _2)=0.\eea
From the set of $\cU _i$, $i=1,2,3,4$ we form four  2-component vectors
\bea
U_1=\begin{pmatrix}
- w_-\\
- w_0+i
\end{pmatrix},\quad U_2=\begin{pmatrix}w_0+i\\-w_+
\end{pmatrix},\\
U_3=\begin{pmatrix}
- w_+\\
- w_0+i
\end{pmatrix},\quad U_4=\begin{pmatrix}w_0+i\\-w_-
\end{pmatrix}.
\eea
These vectors are not orthogonal
\bea
(U_1,U_2)&=&(w_+-w_-)(w_0+i)=\frac{\kappa (s_{12}+i
   a)}{a^2},\\
   (U_3,U_4)&=&(w_--w_+)(w_0+i)=-\frac{\kappa (s_{12}+i
   a)}{a^2}.\eea
Let us remind that
\bea
V_1&=&n_-
\left(
\begin{array}{c}
 +!\frac{w_-(w_0+i)}
  {1+w_0^2
   }\\
 1 \\
\end{array}
\right)\,V_1=n_-
\left(
\begin{array}{c}
 -\frac{w_-(w_0+i)}
  {1+w_0^2
   }\\
 1 \\
\end{array}
\right), \quad (V_1,V_1)=1
\\V_3&=&n_+\left(
\begin{array}{c}
 \frac{w_+
   \left(w_0+i\right)}{1+w_0^2
   } \\
 1 \\
\end{array}
\right), \quad (V_3,V_3)=1\quad (V_1,V_3)=0,
\eea

To get  the orthogonal basis  instead of non-orthogonal one 
\eqref{B50} we take  vectors $\cU_1,\,\cU_2',\,\cU_3$ and $\cU_4'$
\bea
\cU_1=\begin{pmatrix}
  -w_{-}\\
   -w_0\\
   0\\1
\end{pmatrix},\quad \cU '_2=\cJ\cU _1=\begin{pmatrix}
    0\\-1\\-w_-\\-w_0
\end{pmatrix}, \quad\cU _3=\begin{pmatrix}
    -w_+\\-w_0\\0\\1
\end{pmatrix}, \quad \cU '_4=\cJ\cU _3=\begin{pmatrix}
    0\\-1\\-w_+\\-w_0
\end{pmatrix}\nn\\\label{B50new}
\eea
\subsubsection{$\cK$ description of the Bell state}\label{Bellstate}
Qubit basis states can also be combined to form product basis states. A set of qubits taken together is called  a quantum register. For example, two qubits could be represented in a four-dimensional linear vector space spanned by the following product basis states:
\bea
|00\rangle ={\biggl [}{\begin{smallmatrix}1\\0\\0\\0\end{smallmatrix}}{\biggr ]},\quad 
|01\rangle ={\biggl [}{\begin{smallmatrix}0\\1\\0\\0\end{smallmatrix}}{\biggr ]},\quad |10\rangle ={\biggl [}{\begin{smallmatrix}0\\0\\1\\0\end{smallmatrix}}{\biggr ]},\quad |11\rangle ={\biggl [}{\begin{smallmatrix}0\\0\\0\\1\end{smallmatrix}}{\biggr ]}
\eea

A number of qubits taken together is a qubit register. Quantum computers perform calculations by manipulating qubits within a register.

In these notations two entangled qubits is the 
$|\Phi ^{+}\rangle $ Bell state:

\bea
|\Phi ^{+}\rangle=\frac {1}{\sqrt {2}}(|00\rangle +|11\rangle ).
\eea

\end{document}